\documentclass[twocolumn]{aastex631}
\usepackage{graphicx}% Include figure files
\usepackage{dcolumn}% Align table columns on decimal point
\usepackage{bm}% bold math
\usepackage[T1]{fontenc}
\usepackage{mathptmx}
\usepackage{mathrsfs}
\usepackage{xcolor}
\usepackage{hyperref}
\usepackage{physics}
\usepackage{simplewick}
\usepackage{mathrsfs}
\usepackage[bottom]{footmisc}
\usepackage{threeparttable}
\usepackage{appendix}

\graphicspath{{./}{figure/}}
\shorttitle{TDE Overrepresentation in green-valley galaxies}
\shortauthors{Wang et al.}
\graphicspath{{./}{figure/}}
\newcommand{\R}[1]{\textcolor{red}{#1}}
\newcommand{\B}[1]{\textcolor{blue}{#1}}

\begin{document}
\title{An Explanation for Overrepresentation of Tidal Disruption Events in Post-starburst Galaxies} %\footnote{Released on March, 1st, 2021}}
\author[0000-0001-5019-4729]{Mengye Wang}
\affiliation{Department of Astronomy, School of physics, Huazhong University of Science and Technology, Luoyu Road 1037, Wuhan, China}

\author[0000-0001-7192-4874]{Yiqiu Ma$^*$}
\email{* Corresponding author: myqphy@hust.edu.cn}
\affiliation{Department of Astronomy, School of physics, Huazhong University of Science and Technology, Luoyu Road 1037, Wuhan, China}
\affiliation{Center for Gravitational Experiment, School of physics, Huazhong University of Science and Technology, Luoyu Road 1037, Wuhan, China}

\author[0000-0003-4773-4987]{Qingwen Wu$^*$}
\email{* Corresponding author: qwwu@hust.edu.cn} 
\affiliation{Department of Astronomy, School of physics, Huazhong University of Science and Technology, Luoyu Road 1037, Wuhan, China}

\author[0000-0002-7152-3621]{Ning Jiang}
\affiliation{CAS Key laboratory for Research in Galaxies and Cosmology, Department of Astronomy, University of Science and Technology of China, Hefei, 230026, People's Republic of China}
\affiliation{School of Astronomy and Space Sciences, University of Science and Technology of China, Hefei, 230026, People's Republic of China}

\begin{abstract}
Tidal disruption events\,(TDEs) provide a valuable probe in studying the dynamics of stars in the nuclear environments of galaxies. Recent observations show that TDEs are strongly overrepresented in post-starburst or "green valley" galaxies, although the underlying physical mechanism remains unclear.
Considering the possible interaction between stars and active galactic nucleus\,(AGN) disk, the TDE rates can be greatly changed compared to those in quiescent galactic nuclei. In this work, we revisit TDE rates by incorporating an evolving AGN disk within the framework of the "loss cone" theory. We numerically evolve the Fokker-Planck equations by considering the star-disk interactions, in-situ star formation in the unstable region of the outer AGN disk and the evolution of the accretion process for supermassive black holes\,(SMBHs). We find that the TDE rates are enhanced by about two orders of magnitude shortly after the AGN transitions into a non-active stage. During this phase,  the accumulated stars are rapidly scattered into the loss cone due to the disappearance of the inner standard thin disk. Our results provide an explanation for the overrepresentation of TDEs in post-starburst galaxies.

\end{abstract}


\section{Introduction}    \label{sec:intro}  
A tidal disruption event\,(TDE) occurs when a star passes close enough to a supermassive black hole\,(SMBH) with mass $M_{\rm BH}\lesssim 10^8\,M_\odot$. If the star is disrupted by the tidal force, half of the stellar debris will become gravitationally bound and fall back towards the SMBH, which will trigger a short-lived flare of emission\,\citep[e.g.,][]{Rees1988Natur,Evans1989ApJL,Phinney1989,Ulmer1999ApJ}. The matter fallback rate and the light curve theoretically follow $t^{-5/3}$ over the timescale of months to years, which is a typical feature for TDEs\,\citep{Rees1988Natur}. In the past few years, more than 100 TDE candidates have been reported\,\citep[e.g.,][]{French2020SSRv,Hammerstein2023ApJ,Yao2023arXiv} and the number will dramatically increase due to current and upcoming multi-band time-domain surveys\,\citep[][]{Rau2009PASP,Chambers2016arXiv,Shappee2014AAS,Bellm2019PASP,LSST2019ApJ,eROSTA2012arXiv,wfst2023}. 

%It should be noted that TDE occurs rarely in any individual galaxy. 
With the increasing number of observed TDEs, it is possible to estimate the TDE rates, which strongly depend on the environment surrounding the central SMBH\,\citep[typically $\sim 0.1-10\,{\rm pc}$ for galaxies with $M_{\rm BH}\sim 10^{6-8}\,M_\odot$, see][for reviews]{Stone2020SSR,French2020SSRv}, including the number density distribution and the orbital dynamics of stars. Based on the different TDE samples, the TDE rates are typically around $10^{-5}-10^{-4}\,\rm yr^{-1}gal^{-1}$\,\citep[e.g.,][]{Magorrian1999MNRAS,Wangjx2004ApJ,Arcavi2014ApJ,French2016ApJL,van2018ASPC,Stone2020SSR,Yao2023arXiv}, which is in agreement with that estimated from two-body relaxation\,\citep[see][for reviews]{Merritt2013CQGra,Stone2020SSR}. However, several recent studies show that TDE hosts are significantly overrepresented in the post-starburst or E+A galaxies, whose spectra are characterized by a lack of strong emission lines but with strong Balmer absorption features\,\citep[][]{French2017ApJ,Law-Smith2017ApJ,French2020SSRv,Hamerstein2021ApJ,Hammerstein2023ApJ,Yao2023arXiv}. 
Such post-starburst spectra indicate a low current star formation rate and a recent\,($\lesssim 1\,\rm Gyr$) star formation burst\,\citep[but just ended, see][]{Law-Smith2017ApJ,French2020SSRv}. These Balmer-strong galaxies (including post-starburst or E+A galaxies) are commonly found in the green valley, which represents a transitional population between blue star-forming galaxies and red-quenched galaxies\,\citep[see][]{Salim2014SerAJ,Schawinski2014MNRAS}. Accounting for possible selection effects, the TDE rate is estimated to be $\sim25-50$ times overrepresented in this subclass of galaxies\citep[][]{Law-Smith2017ApJ,Hamerstein2021ApJ}, suggesting a close correlation between TDE occurrence and galactic evolution.

Several explanations were proposed to account for the overenhancement in post-starburst galaxies. The gas involved in the galaxy merger loses a substantial fraction of its specific angular momentum due to tidal torques and dynamical friction processes, falling quickly into the center. Therefore, the galaxies may have an overdense central star cluster during this stage, which leads to a shorter timescale of two-body relaxation\,\citep[][]{Stone-Metzger2016MNRAS,Stone-van2016ApJ,French2020ApJ,Chen2023arXiv}. Indeed, high-resolution SDSS and HST imaging showed that the TDE host galaxies have higher central concentrations on pc-kpc scales\,\citep[][]{Law-Smith2017ApJ,Graur2018ApJ,French2020ApJ}. The second possible channel involves the presence of an SMBH binary. Gravitational perturbations from an SMBH companion can significantly modify the orbits of surrounding stars\,\citep[][]{Ivanov2005MNRAS,Chen2011ApJ,Melchor2023arXiv}. The TDE rates could be 2-4 orders of magnitude larger than that estimated for the case of a single SMBH. However, the duration of the phase of enhanced tidal disruption is quite short, which is only $\sim 10^5\,\rm yr$\,\citep[][]{Ivanov2005MNRAS}. It was also proposed that after the merger of the SMBH binary, the gravitational recoil imparted to the merged SMBH fills its loss cone and results in a TDE rate as high as $\sim 0.1\, \rm yr^{-1}$\,\citep[][]{Stone2011MNRAS}. There are also several other dynamical effects for the post-starburst preference, which include radial anisotropies\,\citep[][]{Stone2018AAS}, nuclear triaxiality\,\citep[][]{Merritt2004ApJ} and eccentric nuclear stellar disk\,\citep[][]{Madigan2018ApJ}, where the secular effects can dramatically increase TDE rates. However, the nuclear cluster mass should be relatively small in these scenarios, so as not to give rise to mass precession that quenches coherent secular evolution\,\citep[see][for a review]{Stone2020SSR}.

In this work, we propose an explanation for the overrepresentation of TDEs in post-starburst galaxies by considering the possible interaction between the nuclear stars and AGN disk, where the variation in nuclear stellar density distribution and orbital dynamics within the parsec scale will strongly affect the TDE rates. To achieve high TDE rates, both a high number density and the presence of high eccentric orbits of stars\,(around the loss cone boundary) are necessary. Interestingly, observations indicate that active galactic nuclei\,(AGNs) prefer to reside in hosts that have experienced a recent starburst \,\citep[$\sim$ a few$\times 10^8\,\rm yr$ ago, see][]{Davies2007ApJ,Schawinski2009ApJ,Wild2010MNRAS,Kaviraj2011MNRAS,Hopkins2012MNRAS}. In these starburst galaxies, which could be triggered by galactic mergers\,\citep[][]{Hopkins2010MNRAS,Volonteri2015MNRAS,Liao2023MNRAS}, both the number density distribution and orbital dynamics of stars could be affected by the starburst and the subsequent AGN activity. Thus, the TDE rates are naturally connected to the history of galactic evolution. We calculate TDE rates by solving the Fokker-Planck\,(FP) equations taking into consideration the influence of the accretion disk, which includes a stable inner and a star-forming unstable outer disk surrounding SMBHs\,\citep[][]{Panzhen2021PhysRevD,Wang2023MNRAS}. The in-situ star formation in the disk\,\citep[][]{wang2021accretion,Fan2023ApJ} and the inward migration resulting from star-disk interactions lead to the accumulation of stars with low eccentricity in the nuclear region of galaxies, due to orbital circularization caused by density wave\,\citep[e.g.,][]{Tanaka2004ApJ}. Once the disk dissipates, the stars are re-scattered towards the loss cone boundary, resulting in a rapid increase in TDE rates. 

This paper is organized as follows: we present our main assumptions and numerical method in Section\,\ref{Sec:model}, and show the evolution of TDE rates in Section\,\ref{Sec:results}. We summarize and discuss our results in Section\,\ref{Sec:discussions}.

\begin{figure*}
\centering
\includegraphics[scale=0.27]{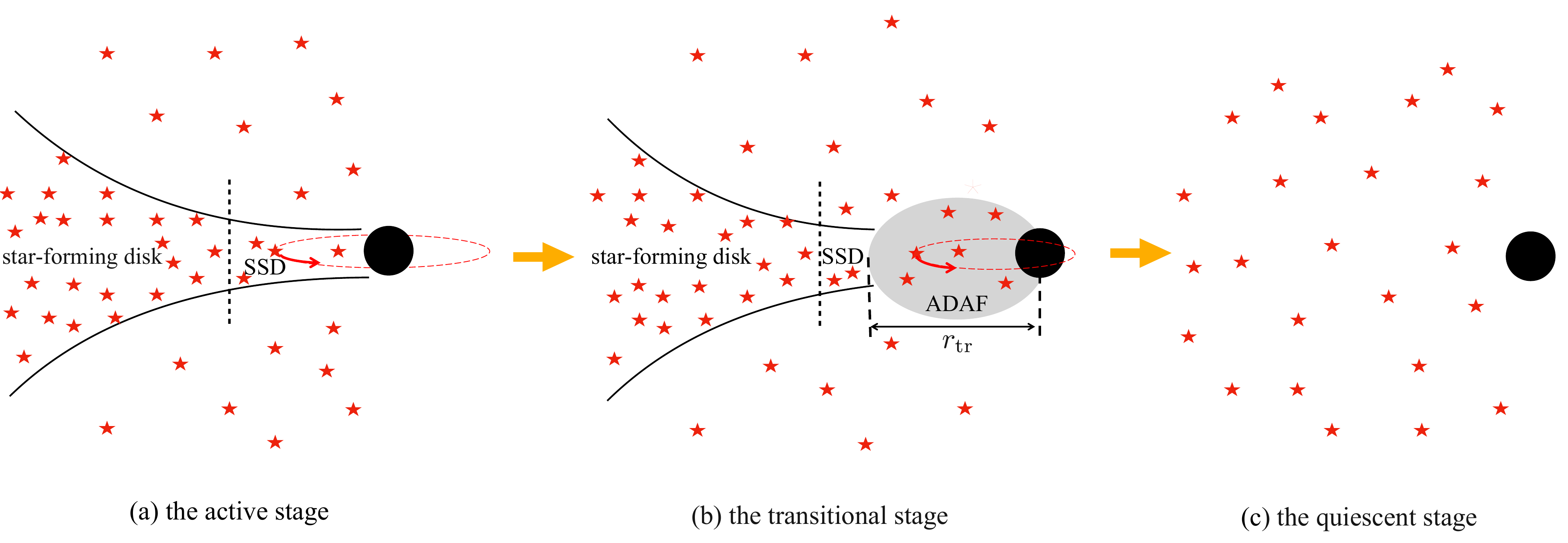}
\caption{
Schematic diagram of accretion disk evolution: the supermassive black hole\,(SMBH) resides at the galaxy's center, accompanied by a surrounding accretion disk and a population of stars. From left to right, Figures\,(a), (b), and (c) represent the SMBH stay in the active stage, transitional stage and quiescent stage, respectively. The red dashed lines in Figure\,(a) and (b) show the star in the circular orbit and eccentric orbit, respectively.
The active stage (a): apart from the stars staying outside of the AGN disk, many stars are formed in the outer star-forming disk or captured by the disk. The stars embedded in the AGN disk normally stay in circular orbit as the disk gas, which are far from the loss cone boundary. 
The transitional stage (b): the standard thin disk\,\citep[Shakura-Sunyaev disk, or SSD, see][]{Shakura1973A&A} gradually switches into an optically thin, geometrically thick advection-dominated accretion flow\,(ADAF) within a truncation radius($r_{\rm tr}$) after the accretion rate is lower than a critical value. The disk-component stars can be quickly scattered into the loss cone boundary through two-body relaxation and lead to a rapid increase in the TDE rates.  The quiescent stage(c): the galaxy enters the quiescent stage at a very low accretion rate, where the SSD would fully disappear. }
\label{fig:0}
\end{figure*}

\section{Model} \label{Sec:model}
The TDE rates can be calculated by integrating the flux in the direction of angular momentum at the boundary of the loss cone, which strongly depends on the environment surrounding the central SMBH and varies significantly during different stages of galactic evolution. The presence or absence of the accretion disk will significantly affect star formation, stellar evolution, and stellar orbits in the nuclear region. However, the precise evolution of the accretion process in the galactic center remains unclear. Following the positive correlation between the accretion rate of the central SMBH and the star formation rate\,(SFR)  of galaxies given by simulations and observations, we simply assume a fast rise and slow decay for the evolution of accretion rate \,\citep{Behroozi2013ApJ,Carnall2018MNRAS}, which is described by a double-power-law form:
\begin{equation}  \label{eq:BHAR}
    \dot{M}(t) = \dot{M}_{0} \left[\left(\frac{t}{\tau}\right)^a + \left(\frac{t}{\tau}\right)^{-b}\right]^{-1}
\end{equation}
where $a$ and $b$ are the falling and rising slopes, respectively, and $\tau$ is related to the time at which the accretion rate peaks. As shown in Figure\,\ref{fig:0}, we divide the galaxy into three stages according to the accretion rate: the active stage, the transitional stage and the quiescent stage.

\subsection{The AGN disk and the star-disk interactions}

The accretion disk is normally optically thick and geometrically thin in bright AGNs\,\citep[][]{Shakura1973A&A}. The standard thin disk is generally gravitationally unstable, leading to the collapse of self-gravitating clouds and forming stars at distances greater than several thousand Schwarzschild radii\,\citep[][]{Toomre1964ApJ}. In this work, we adopt the TQM disk in the active stage\,\citep[see][]{Thompson2005ApJ}, which consists of an inner standard thin disk and an outer disk undergoing star formation. The star formation rate in the unstable region of the TQM disk, $\dot{M}_\star$, can be estimated from the well-known Kennicutt–Schmidt law\,\citep[$\dot{\Sigma}_{\dot{M}_\star}\sim 0.017\Sigma_{\rm gas}\Omega_{\rm gas}$, see][]{K-S-law1998ApJ}. In this work, we follow \cite{Dittmann2020MNRAS} to obtain the $\dot{M}_\star$, which is determined by the viscosity parameter $\alpha$ and the efficiency parameter $\epsilon$\,\citep[i.e., the rest mass energy from star formation is converted into radiation, see][for in detail]{Dittmann2020MNRAS}. %\,(see equation \ref{eq:source term}).

In low-luminosity AGNs, the standard thin disk will switch into an optically thin, geometrically thick advection-dominated accretion flow\,(ADAF) when the accretion rate $\dot{m}$\,($\equiv \dot{M}/\dot{M}_{\rm Edd}$) falls below the critical value $\dot{m}_{\rm crit,1}$\,\citep[typically $\dot{m}_{\rm crit,1}\sim 0.01$, see][]{Narayan1994ApJ}. The physical mechanism driving the disk transition is still unclear, but the evaporation/condensation processes may play a key role\,\citep[e.g.,][]{Qiao2009PASJ}. In this work, we adopt a truncation radius of $r_{\rm tr} = 17.3 \dot{m}^{-0.886}\alpha^{0.07}$\,\citep[in the unit of Schwarzschild radius $r_s$,][]{Taam2012ApJ} to represent the transition between these two accretion modes. 
Once the standard thin disk transitions to an ADAF, we neglect star-disk interactions due to the low gas density in the hot ADAF region. We assume that the standard disk is completely converted to the ADAF when the accretion rate $\dot{m}<\dot{m}_{\rm crit,2}\sim 0.001$. Since the truncation radius $r_{\rm tr}\sim 10^4\,r_s$ at this critical accretion rate.

In the AGN disk, the migration timescale and eccentricity damping timescale caused by the density wave excitation are well described by the following expression\,\citep[][]{Cresswell2008A&A}:
\begin{equation} \label{eq:tau_mig}
\begin{split}
 &   \tau_{\rm mig} =\frac{\tau_{\rm wave}}{h^{2}}\frac{1+(e/2.25h)^{1.2}+(e/2.84h)^6}{1-(e/2.02h)^4},\\
 &    \tau_e = \frac{\tau_{\rm wave}}{0.78}\left[ 1-0.14(e/h)^2+0.06(e/h)^3  \right]
\end{split}
\end{equation}
with
\begin{equation}
    \tau_{\rm wave} = \frac{M}{m_s}\frac{M}{\Sigma a^2} \frac{h^4}{\Omega}
\end{equation}
where $m_s$ is the mass of stars, $M=M(<a)$ is the total mass consisting of the SMBH, stars and disk gas within the radius $a$, $e$, $a$ and $\Omega$ are the orbital eccentricity, semi-major axis and angular velocity of stars, $\Sigma$ and $h=H/a$ are the disk surface density and the disk aspect ratio, respectively.

\subsection{FP equations and TDE rates} \label{sec:2.2}

In this section, we introduce the orbit-averaged FP equations and the numerical set-up, which is mainly based on the work of \cite{Cohn1978ApJ,Stone-Metzger2016MNRAS,Panzhen2021PhysRevD,Broggi2022MNRAS,Wang2023MNRAS}, where the numerical method for the FP equations can be found in \cite{Panzhen2021PhysRevD,Broggi2022MNRAS,Wang2023MNRAS}. We consider a central SMBH with mass $M_{\rm BH}$ surrounded by a spherically stellar cluster with total mass $M_s$ and a star-forming disk. The stars in the stellar cluster are assumed to have a single mass component $m_s$. The initial stars' distribution follows Tremaine's cluster model\,\citep{Tremaine1994}, which is given by 
\begin{equation}
    n_{s}(r)=\frac{M_s}{m_s}\frac{3-\gamma}{4\pi}\frac{r_a}{r^\gamma (r+r_a)^{4-\gamma}},
\end{equation}
where $r_a=4GM_{\rm BH}/\sigma_*^2\equiv 4r_h$ is the radius of density transition\,($r_h$ is the influential radius of the SMBH with mass $M_{\rm BH}$), $\sigma_*$ is the stellar velocity dispersion given by $M_{\rm BH}-\sigma_\star$ relaxation \citep[][]{Tremaine2002,Gultekin2009},
\begin{equation} \label{M_sigma_relaxation}
M_{\rm BH}=1.53\times 10^6M_\odot \left(\frac{\sigma_*}{70 {\rm km}/s}\right)^{4.24}
\end{equation}
and $\gamma\sim 1.2-1.8$ is the density scaling power index \citep[][]{Tremaine1994,Binney2008}. With the above density profiles, the collective gravitational potential background can be derived as:
\begin{equation} \label{eq:phi}
\phi (r)=\frac{GM_{\rm BH}}{r} +\frac{GM_s}{r_a}\frac{1}{2-\gamma}\left[1-\left(\frac{r}{r+r_a}\right)^{2-\gamma}\right].
\end{equation}

Following \cite{Cohn1978ApJ}, we define the specific orbit energy and dimensionless specific orbital angular momentum: 
\begin{equation}
    E\equiv \phi(r)-\frac{v^2}{2}, \quad R\equiv \frac{J^2}{J_c^2(E)}
\end{equation}
where $J$ is the specific orbital angular momentum and $J_c(E)$ is the specific orbital angular momentum of a star with specific energy $E$ on a circular orbit.
In this case, the initial distribution function in the $(E,R)$-phase space is given by %only depending on the energy $E$ and the space number density of stars/sBHs by
\begin{equation}
f_i(t=0,E,R)=\frac{\sqrt{2}}{(2\pi)^2}\frac{{\rm d}}{{\rm d}E}\int^E_0\frac{{\rm d}n_{s}}{{\rm d} \phi}\frac{{\rm d} \phi}{\sqrt{E-\phi}},
\end{equation}
where $n_{s}$ is the number density of stars\,\citep[][]{Tremaine1994,Binney2008}.

Given initial distributions of stars, $f_i(t=0,E,R)$, their evolution is governed by the orbit-averaged Fokker-Planck equation\,\citep[][]{Cohn1978ApJ}
\begin{equation}
    C\frac{\partial f_i}{\partial t} = -\frac{\partial }{\partial E}F_E - \frac{\partial }{\partial R}F_R,
\end{equation}
where $C = C(E, \,R)\equiv 4\pi P(E,R)J_c^2(E)$ is a normalization coefficient, $P(E,R)$ is the orbital period and $F_{E,R}$ is the flux in the $E/R$ direction
\begin{equation} 
\begin{split}
&F_E=-\left(D_{EE}\frac{\partial f_i}{\partial E}+D_{ER}\frac{\partial f_i}{\partial R}+D_E\, f_i\right),\\
&F_R=-\left(D_{RR}\frac{\partial f_i}{\partial R}+D_{ER}\frac{\partial f_i}{\partial E}+D_R \,f_i\right),\\
\end{split}
\end{equation}
with the diffusion coefficients $\{D_{EE},\,D_{RR},\,D_{ER}\}$ and the advection coefficients $\{D_E,\,D_R\}$ as functions of $f_i(t,E,R)$\,\citep[which is defined in][]{Cohn1978ApJ}. We evolve the $f_i(t=0,E,R)$ to a stable state $f_i(t=t_f,E,R)$ according to the above FP equations and set the stable state as the initial state for subsequent calculations.

To calculate the TDE rates, we first obtain the evolution of the stars' distribution function by solving the FP equations above the three stages of galaxies:

In the active stage, the stars can be divided into two populations: the disk component (stars embedded within the TQM disk) and the cluster component (stars stay outside of the TQM disk). Initially, we set the distribution function of cluster-component stars $f(t=0,E,R)=0.99\, f_i(t=t_f,E,R)$ and disk-component stars $g(t=0,E,R)=0.01\, f_i(t=t_f,E,R)$.
The evolution of cluster- and disk-component distribution functions is governed by the orbit-averaged FP equations\,\citep[][]{Panzhen2021PhysRevD,Wang2023MNRAS}
\begin{equation}
\begin{split}
 &   C\frac{\partial f}{\partial t} = -\frac{\partial}{\partial E}F_E^f - \frac{\partial}{\partial R}F_R^f - S_{\rm cap},\\
&   C\frac{\partial g}{\partial t} = -\frac{\partial}{\partial E}F_E^g - \frac{\partial}{\partial R}F_R^g + S_{\rm cap} + S_{\rm dSF},
\end{split}
\end{equation}
 where $F_{E,R}^X$ is the flux in the $E/R$ direction
\begin{equation} \label{FE_FR}
\begin{split}
&F_E^X=-\left(D_{EE}\frac{\partial X}{\partial E}+D_{ER}\frac{\partial X}{\partial R}+D_E^X\, X\right),\\
&F_R^X=-\left(D_{RR}\frac{\partial X}{\partial R}+D_{ER}\frac{\partial X}{\partial E}+D_R^X \,X\right),\\
\end{split}
\end{equation}
\\
with $X=\{f,\,g\}$, the diffusion coefficients $\{D_{EE},\,D_{RR},\,D_{ER}\}$, and the advection coefficients $\{D_E^X,\,D_R^X\}$ incorporating two-body scatterings and star-disk interactions. The advection coefficients in the disk are modified as 
\begin{equation}
\begin{split}
&D_E^g = D_E - C\frac{E}{\tau_{\rm mig}},\\
&D_R^g = D_R - C\frac{1-R}{\tau_{\rm e}},\\
\end{split}
\end{equation}
where $D_E$ and $D_R$ are the advection terms in the quiescent stage and $\tau_{\rm mig}$, $\tau_e$ are given by equation\,\eqref{eq:tau_mig} . It should be noted that the star-disk interactions will dominate the evolution of distribution function in the cluster only for the low inclination orbiters\,($\theta\lesssim\pi/6$) since the migration torque by density wave decays quickly as the orbital inclination increase\,\citep[e.g.,][]{Cresswell2008A&A}. Thus, we ignore the effect of the star-disk interactions on the cluster-component stars, which means that the rate of TDEs in the active stage could be close to that in the quiescent stage\,\citep[see also,][]{Macleod2020ApJ,Teboul2022arXiv}.

The source term $S_{\rm cap}$ arises from stars captured by the disk and $S_{\rm dSF}$ arises from in-situ star formation in the outer unstable region of the disk, with 
\begin{equation}  \label{eq:source term}
    S_{\rm cap }(E,R) = \mu_{\rm cap} C\frac{f}{\tau_{\rm mig}},  \quad  S_{\rm dSF} (E, R)= \frac{1}{m_{s}}\frac{\partial \,\rm \dot{M}_\star}{\partial E\,\partial R},  
\end{equation}
where $\mu_{\rm cap}=0.01 $ is a parameter quantifying the disk capture efficiency\,\citep[see][]{Panzhen2021PhysRevD}, $\tau_{\rm mig}$ is the migration timescale by star-disk interactions, $\dot{M}_\star (E,R) = \dot{M}_\star(E)\,\delta(R-1)$ and the $\dot{M}_\star(E)$ is obtained from the TQM disk model\,\citep[][]{Dittmann2020MNRAS}. Here, we simply assume that the stars formed in the TQM disk are all in circular orbits, i.e., $R=1$\,\citep[][]{Derdzinski2022arXiv}. For numerical simplification, we adopt the phenomenological treatment\,($S_{\rm cap}$) for the disk capture process rather than calculate the change in orbital inclination\,($\theta$) of the stars' distribution function by solving the FP equations with the diffusion\,($D_{\theta\theta}$) and advection\,($D_\theta$) terms\,\citep[see][]{Pan2022PhRvD}.

In the transitional stage, the inner thin disk will be gradually evaporated within a truncation radius\,($r_{\rm tr}$), and we neglect the star-disk interactions for the region within $r<r_{\rm tr}$. When the standard accretion disk disappears completely, the galaxy will enter into the quiescent stage, where the advection term $D_E^g$ and $D_R^g$ are substituted by $D_E$ and $D_R$ in the equation\,\eqref{FE_FR}\,\citep[see][]{Cohn1978ApJ,Panzhen2021PhysRevD,Broggi2022MNRAS,Wang2023MNRAS}. Here, we do not include the process of the scattering of the disk-component stars into the cluster after the disk disappears. This process can be described by the diffusion\,($D_{\rm \theta\theta}$) and advection coefficients\,($D_\theta$) in the FP equations. However, both the diffusion coefficients $D_{E\theta}$ and $D_{R\theta}$ are equal to zero\,\citep[see the appendix B in][and reference therein]{Pan2022PhRvD}, which means that the values of energy and angular momentum do not change as the stars' inclination changes in this process. Therefore, the $D_{E\theta}, D_{R\theta}$ will not affect the TDE rates.

In the active stage, we set $F_R=0$ in the loss cone boundary due to the short eccentricity decay timescale of star-disk interactions\,\citep[$\tau_{\rm wave}/\tau_{\rm disk}\sim 10^{-5}$, where $\tau_{\rm disk}$ is the duty cycle of AGN, see][]{Panzhen2021PhysRevD}. Furthermore, the distribution function of stars in the quiescent stage is set as the initial conditions of the active stage in our simulation. In the transitional stage with intermediate accretion rate, we adopt $F_R=0$ for $r>r_{\rm tr}$ and the standard loss cone boundary conditions for $r\leq r_{\rm tr}$\,\citep[see][]{Cohn1978ApJ}. In the quiescent stage, we still adopt the standard loss cone boundary conditions. The TDE rates are determined by the flux in the $R$-direction:
\begin{equation}
    \Gamma_{\rm TDE} (M_{\rm BH},t) = \int -(F_R^g+F_R^f) \,dE.
\end{equation}

\begin{figure*}
\gridline{\fig{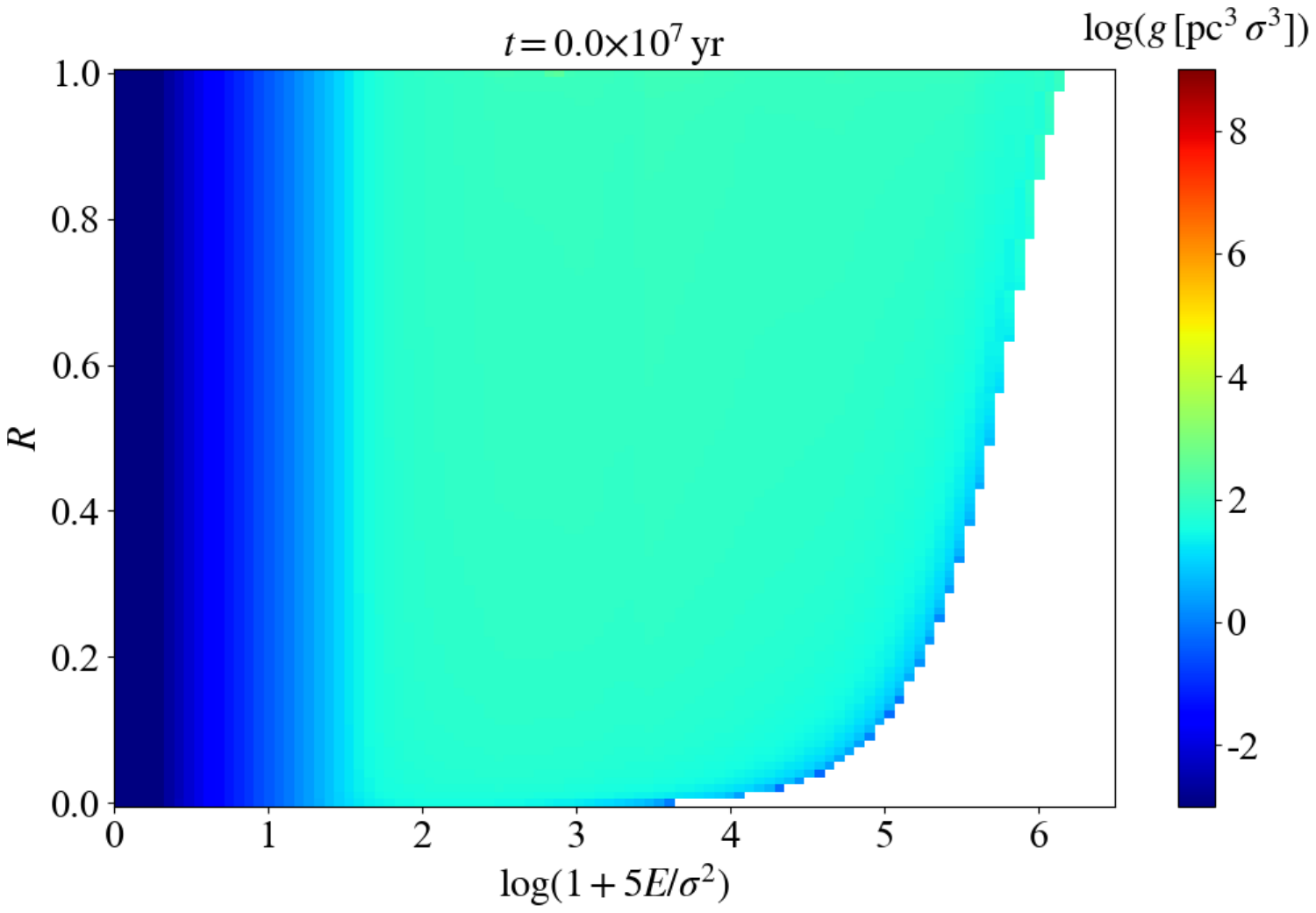}{0.5\textwidth}{(a)}
          \fig{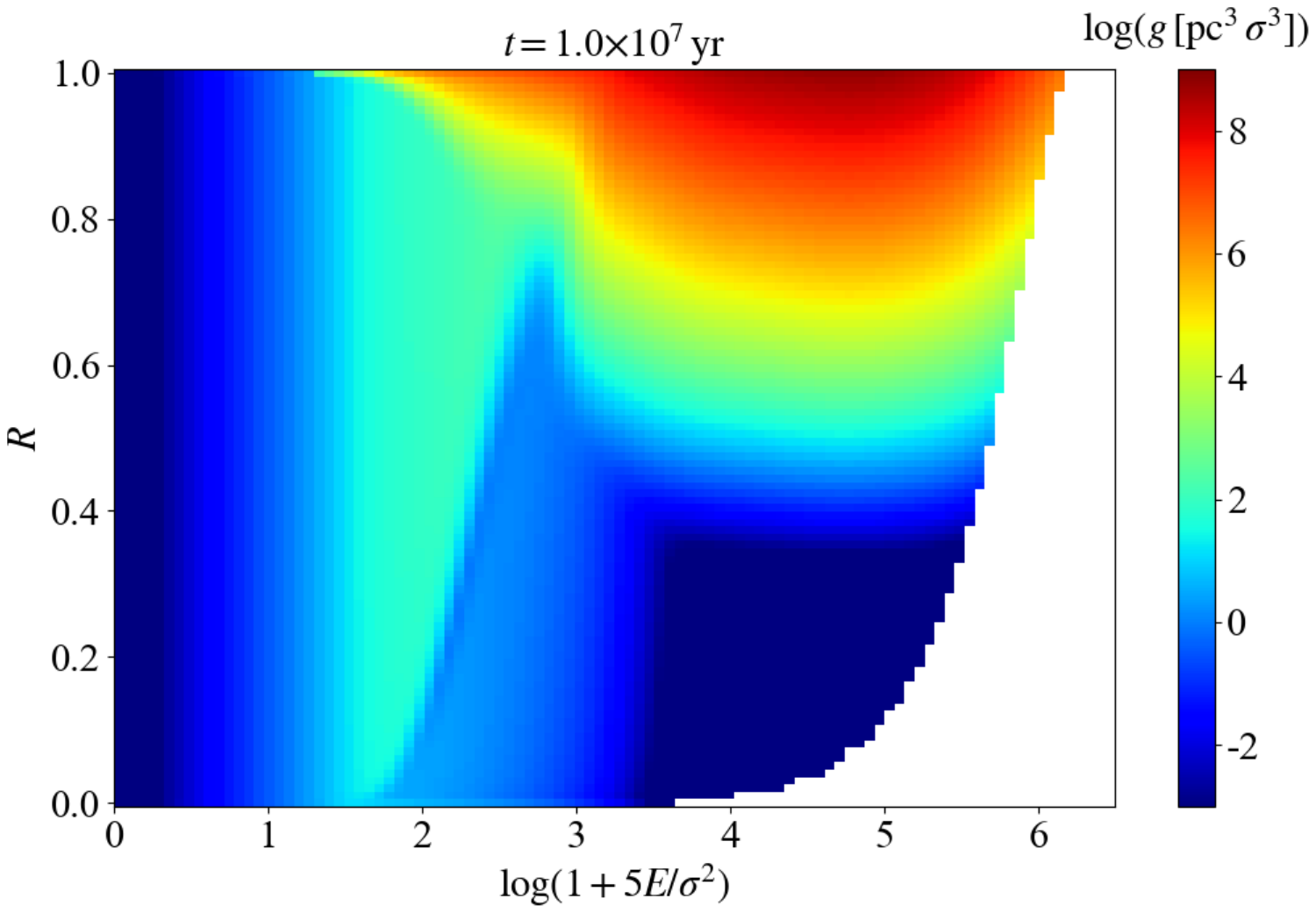}{0.5\textwidth}{(b)}
          }
\gridline{
          \fig{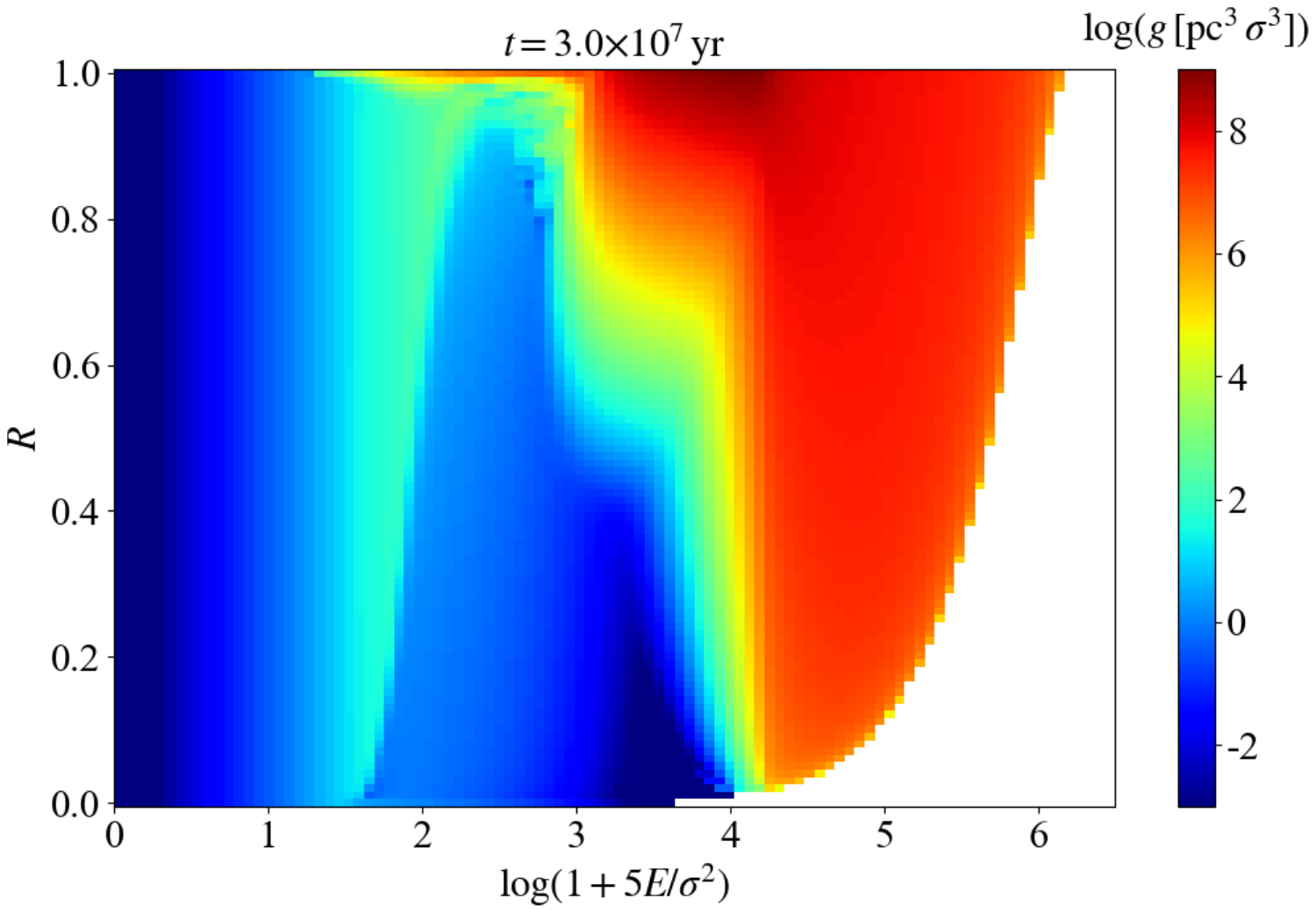}{0.5\textwidth}{(c)}
          \fig{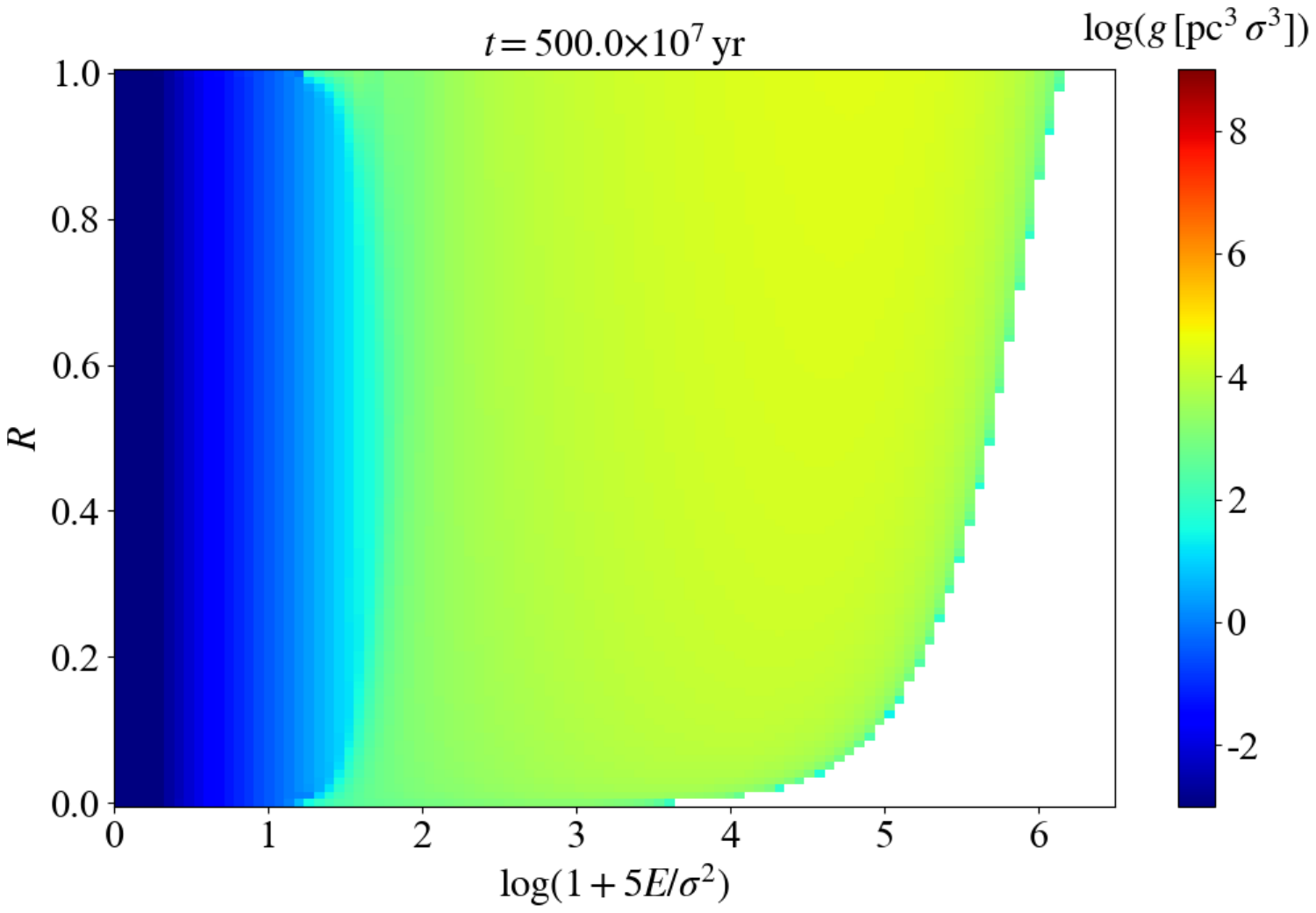}{0.5\textwidth}{(d)}
          }
\caption{The distribution function of the disk-component stars in the $(E,\,R)$ space, $g(t,E,R)$, in the unit of ${\rm pc}^{-3} \cdot \sigma^{-3}$ with SMBH mass $M_{\rm BH} =10^{6.75}\,M_\odot$. (a), (b), (c) and (d) show the distribution function in the initial stage, active stage\,($\dot{m}=0.06$), transitional stage\,($\dot{m}=0.009$) and the quiescent stage according to the different accretion rate, respectively. The stars will gather at the $R=1$ boundary in the ($E$, $R$) space in the active stage, while they will be scattered to the loss cone boundary in the transitional stage. }
\label{fig:1}
\end{figure*}

\section{results} \label{Sec:results}
Following the above procedure, we consider a model consisting of stars orbiting around an SMBH, where part of the stars are embedded in the accretion disk while others still stay in a spherical star cluster. For simplicity, we set a monochromatic star mass distribution  $m_s=M_\odot$. We assume that the density power-law index of $\gamma=1.5$\,\citep[$\rho\propto r^{-\gamma}$, see][for details]{Panzhen2021PhysRevD,Wang2023MNRAS}. 
%For accretion disk, We adopt the $\alpha$-viscosity parameter $\alpha = 0.1$ and the efficiency parameter $\epsilon=0.4$ (???). 
In this work, we assume $\dot{M}_{0}=\dot{M}_{\rm Edd}$, $a=1.7$, $b=4.0$, $\tau=2\times 10^6\,$yr, which will not affect the main result of overrepresentation of TDEs in the transitional stage. 

\begin{figure}
\centering
\includegraphics[scale=0.35]{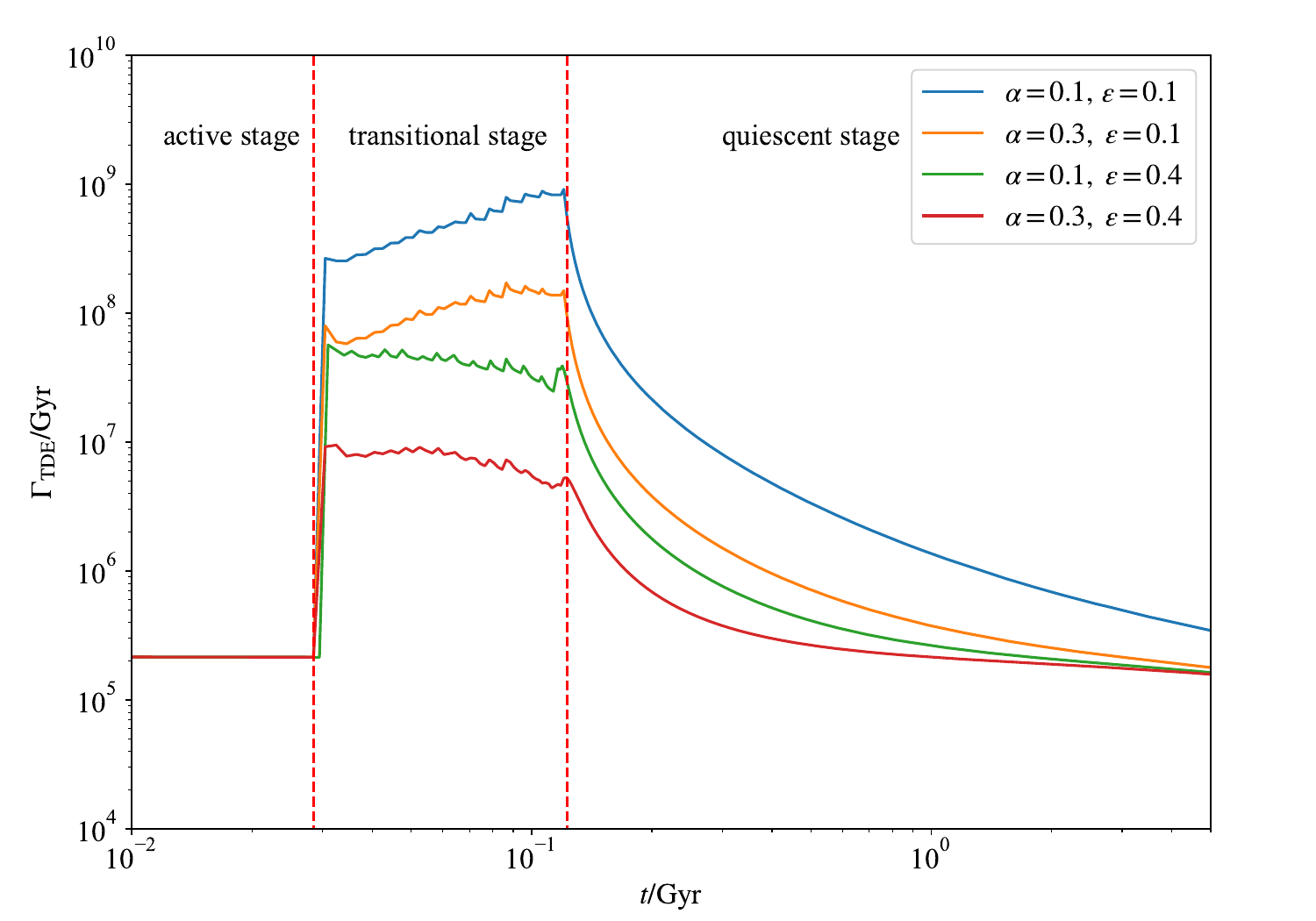}
\caption{Evolution of TDE rates over time with $M_{\rm BH}=10^{6.75}\,M_\odot$. The blue, orange, green and red show the evolution of TDE rates with the different disk parameters\,\citep[ $\alpha=0.1,0.3$ and $\epsilon=0.1,0.4$, which is adapted from][]{Dittmann2020MNRAS}, respectively. The two red dotted lines divide the evolutionary stage into three parts: the active stage, the transitional stage and the quiescent stage.}
\label{fig:2}
\end{figure}

\subsection{The simulation results of the FP equations}
In Figure\,\ref{fig:1}, we show the stars' distribution function $f(t,E,R)$ with of the different evolution stage in $(E,R)$ space in the unit of ${\rm pc}^{-3}\cdot\sigma^{-3}$ with the SMBH mass $M_{\rm BH}=10^{6.75}\,M_\odot$. In Figure\,\ref{fig:1}(a), we show the steady state of the evolution in the quiescent stage as the initial conditions of the active stage, where we assume that the accretion disk will present in some stage of galaxy evolution or duty-cycle effect. The distribution functions of active, transitional and quiescent stages are shown in Figure\,\ref{fig:1}(b), (c) and (d), respectively.
In the active stage, star-disk interactions (e.g., migration and damping of eccentricity) and star formation in the disk will dominate the evolution of the stars' distribution. As shown in Figure\,\ref{fig:1}(b) with the accretion rate $\dot{m}=0.06$, stars will gather at the boundary of $R=1$ in the ($E,\,R$) space due to the circularizing induced by the density wave excitation. The stars in the circular orbits are far from the loss cone boundary, which hardly contributes to the TDE rates.  In the transitional stage shown in Figure\,\ref{fig:1}(c) with accretion rate $\dot{m}=0.009$, the stars will be excited to the high eccentricity under the two-body scattering and enter the loss cone boundary as the star-disk interactions and star formation gradually vanish. In the quiescent stage, the distribution of stars will gradually return to the initial condition\,(see Figure\,\ref{fig:1}d). 

\subsection{Evolution of the TDE rates}
We present the evolution of TDE rates with different disk parameters\,($\alpha,\,\epsilon$) in Figure\,\ref{fig:2}. The green, orange, blue and red lines show the disk parameters $\alpha = 0.1$, $0.3$ and $\epsilon = 0.1$, $0.4$ with $M_{\rm BH} = 10^{6.75}\,M_\odot$, respectively. The two red dotted lines divide the evolutionary stage into the active stage, transitional stage and quiescent stage. The active, transitional and quiescent stages are divided by the two red dotted lines. In the active stage, the low-inclination stars would fall into accretion disk and be in circular orbits due to star-disk interactions within a timescale of $10^{2-4}\,\rm yr$\,\citep[e.g.,][]{Tremaine2002,Panzhen2021PhysRevD} , while the high-inclination stars remain in the cluster. Therefore, the TDE rates in the active stage are mainly caused by the scattering of stars in high-inclination orbits, which are close to that in the quiescent galaxies as shown in Figure\,\ref{fig:2}. During the transitional stage, when the inner standard thin disk disappearances, many previously embedded stars within the accretion disk are scattered, leading to a rapid increase in TDE rates. We find that the TDE rates during this stage are about $10^{-2}-10^{-1}\,\rm yr^{-1}gal^{-1}$, which are enhanced by 2-3 orders of magnitude compared to that in the normal quiescent galaxies. After the transitional stage, the TDE rates gradually decrease and return to levels similar to those in quiescent galaxies after about $1\,{\rm Gyr}$. For the four different sets of disk parameters, the average TDE rates over 1 Gyr are about 320\,(65, 22 and 5) times higher than those in normal quiescent galaxies for the green\,(orange, blue and red) lines as shown in Figure\,\ref{fig:2}, which is consistent with the overrepresentation of TDEs in post-starburst galaxies\,\citep[e.g.,][]{French2020ApJ}.

\begin{figure}
%\centering
\hspace*{-30pt}
\raggedright
\includegraphics[scale=0.35]{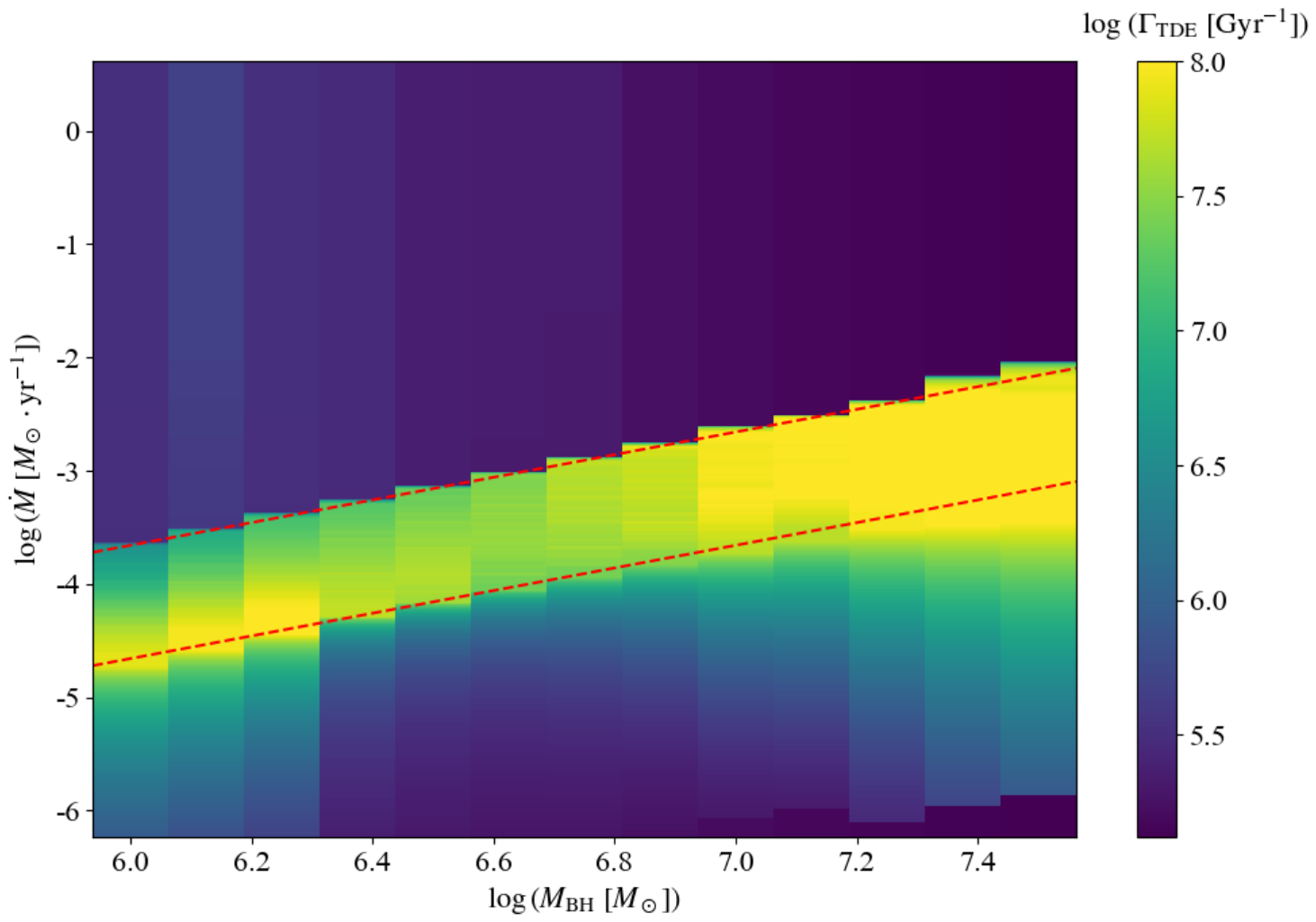}
\caption{The TDE rates distribution in ($M_{\rm BH}$, $\dot{M}$) space with $\alpha=0.1$ and $\epsilon=0.4$. The red dashed lines show the transition accretion rate\,($\dot{m}_{\rm crit,2}<\dot{m}\left(\equiv \dot{M}/\dot{M}_{\rm Edd}\right)<\dot{m}_{\rm crit,1}$, here, we assume $\dot{m}_{\rm crit,1}=0.01$ and $\dot{m}_{\rm crit,2}=0.001$). }
\label{fig:3}
\end{figure}

We also present the distribution of the TDE rates in phase space of ($M_{\rm BH}$, $\dot{M}$) with $\alpha = 0.1$, $\epsilon = 0.4$ in Figure\,\ref{fig:3}. The red lines indicate the transitional region between active and non-active galaxies\,($\dot{m}_{\rm crit,2}<\dot{m}<\dot{m}_{\rm crit,1}$). Utilizing the empirical relaxation between the accretion rate and SFR found in observations, ${\rm log}\, \left( {\rm SFR}/(M_\odot \cdot {\rm yr}^{-1})\right) = 0.73 \,{\rm log} \,\left( \dot{M}/(M_\odot \cdot {\rm yr}^{-1})\right) + 1.55$\,\citep[e.g.,][]{Satyapal2005ApJ,Wu2006PASP,Hopkins2010MNRAS}, we estimate the SFR corresponding to the accretion rate of the transitional region: 
\begin{equation}
\begin{split}
&-11.56 +0.18\,{\rm log} \left(\frac{M_\star}{10^{10}\,M_\odot}\right) < {\rm log} \left(\frac{{\rm SFR}}{M_\star\cdot \rm yr^{-1}} \right) \\
& \quad\quad\quad < -10.83 +0.18\,{\rm log} \left(\frac{M_\star}{10^{10}\,M_\odot}\right),
\end{split}
\end{equation}
where the relaxation ${\rm log} \left(M_{\rm BH}/10^9M_\odot \right) = - 1.83 + 1.62\, {\rm log} \left( M_\star/3\times 10^{10}M_\odot\right)$ is adopted \citep[$M_\star$ is the total stellar mass, see][]{Yao2023arXiv}. We find that the SFR corresponding to the transitional stage is roughly consistent with that estimated for the galaxies in the green valley\,\citep[$-11.8<{\rm log} \left({\rm SFR}/(M_\star\cdot \rm yr^{-1}) \right)<-10.8$, see][]{Salim2014SerAJ}.

\section{Discussions}\label{Sec:discussions}

In this work, we explore the TDE rates based on a modified loss cone theory, where both star-disk interactions and star formation in the outer disk are considered. We find that the star formation in the outer unstable region of the AGN disk and the star-disk interactions will lead to the accumulation of stars in the nuclear region of galaxies. As the orbital circularization induced by the density wave disappears in the low-luminosity AGNs and quiescent galaxies, the stars will be scattered to the loss cone boundary and lead to a rapid increase in the TDE rates, which offers a possible explanation for the overrepresentation of TDE rates observed in green valley (or post-starburst) galaxies\,\citep[][]{French2016ApJL,French2017ApJ,Law-Smith2017ApJ,French2020SSRv,Hamerstein2021ApJ,Hammerstein2023ApJ,Yao2023arXiv}.

Considering the in-situ formation and capture effect, it is not surprising for the presence of overdense nuclear star clusters and compact remnants in the AGN disk. The disk stars experience much different interactions than the stars in a vacuum, where the disk stars could suffer star-disk interactions, accretion, etc. The possible gravitational wave event associated with EM counterparts of GW190521 from the merger of two BHs is suggested to be the possible candidate that formed in AGN disk\,\citep[][]{Ashton2021CQGra}. Furthermore, it is suggested that several AGNs may be super-solar metallicity based on the line ratios from the broad lines, which may be enriched by the supernovae due to the disk stars\,\citep[e.g.,][]{Toyouchi2022MNRAS,Fan2023ApJ}. The TDE rates are closely correlated with the number density of stars in the nuclear region, where the nuclear star clusters may play a key role in connecting the SMBH growth and properties of host galaxies. Many observations and theories showed a strong correlation between starburst and AGN activity\,\citep[e.g.,][]{Brotherton_1999,Satyapal2005ApJ,Wu2006PASP,Kaviraj2011MNRAS,Dai2018MNRAS}, which could be triggered by the galactic mergers\,\citep[][]{Hopkins2010MNRAS,Volonteri2015MNRAS,Liao2023MNRAS}. The starburst, nuclear star cluster and SMBH activity in the host galaxy should be closely correlated, even though there is a possible delay\,(about a few $10^8\,\rm yr$) between the peak of star formation and AGN activity\,\citep[][]{Davies2007ApJ,Schawinski2009ApJ,Wild2010MNRAS,Hopkins2012MNRAS,Marvin2016MNRAS}. Future high-resolution optical/UV observations on the nuclear star clusters with different nuclear SMBH activities can further shed light on this issue.

However, to explain the TDE rates overenhancement in green valley\,(or post-starburst) galaxies, we have adopted several assumptions, which are worth further discussion:

$\bullet$ Due to the complexity of black hole accretion and the limited observational data available, the evolution of the accretion rate and the disk model are quite uncertain. We assume that the accretion rate evolves as a double-pow-law form and divide AGN evolution into three stages: active stage\,($\dot{m}>\dot{m}_{\rm crit,1}$), transitional stage\,($\dot{m}_{\rm crit,2}\leq\dot{m}\leq \dot{m}_{\rm crit,1}$) and quiescent stage\,($\dot{m}<\dot{m}_{\rm crit,2}$). The transitional accretion rate could greatly affect the time at which the peak of the TDE rates occur, while the disk model parameters\,\citep[e.g., $\alpha$ and $\epsilon$, see][]{Dittmann2020MNRAS} could affect the magnitude of the peak of the TDE rates. However, the uncertainties of the disk model and the accretion rate's evolution will not affect the main conclusion that the enhancement of TDE rates will happen just after the disk transition.

$\bullet$ Though the correlation between starburst and AGN activity has been suggested by many authors, the quantitative relationship between accretion rate and SFR is still unknown. It is difficult to confirm whether the transitional stage of the AGN corresponds to the stage of green valley\,(or post-starburst) galaxies. 

$\bullet$ The spherical symmetry of the stars' distribution will be broken by the star-forming disk, which could affect the background gravitational potential $\phi(r)$. However, the peak of star formation rate in the disk is about $0.01\,M_\odot \cdot {\rm yr}^{-1}$ for SMBH mass $10^{6.75}\,M_\odot$, which is located at $10^5$-$10^6 \,r_g$. In this region, the background gravitational potential is still dominated by the central SMBH and the original spherical symmetry star cluster where the gravitational potential asymmetry caused by the accretion disk and disk-component stars is less than ten percent. Therefore, we still assume that the background gravitational potential $\phi(r)$ is a constant function as shown in equation\,\eqref{eq:phi}, which is spherically symmetric and stable.

$\bullet$ The Kozai-Lidov effect\,\citep[][]{Lidov1962,Kozai1962AJ} could be triggered by the accretion disk\,\citep[][]{Karas2007A&A}, which can cause a secular oscillation of the cluster-component stars' inclination and eccentricity. The analytical and simulated studies have shown that the Kozai-Lidov effect could increase TDE rates by several times for the disk with mass $M_d\sim 0.01M_{\rm BH}$\,\citep{Karas2007A&A,Terquem2010MNRAS,Kennedy2016MNRAS}, which means that the TDE rate could be higher than that in the active stage as shown in Figure\,\ref{fig:2}. Note that the origin of large amplitude AGN variability is poorly understood, which could make it difficult to distinguish it from TDEs\,\citep[see][for a review]{van2020SSRv}. Up to now, only a few TDE candidates in AGNs have been found, most of which were discovered in low-luminosity galaxies\,\citep[e.g.,][]{Blanchard2017ApJ,Yan2018MNRAS,Liu2020ApJ,Homan2023A&A}. In this work, we neglect the Kozai-Lidov effect since we do not aim at the TDE rate in the active stage. In the transitional stage, the number of the disk-component stars is about 2-3 orders of magnitude higher than that of the cluster-component stars within the region where the peak of star formation rate in the disk is located\,($\sim 10^5-10^6\,r_g$). However, for stars with a low\,(or zero) inclination angle with respect to the disk, the Kozai-Lidov process can not be efficiently excited\,\citep[e.g.,][]{Naoz2013MNRAS}. Compared to the TDE rate enhancement by in-situ star formation, TDE rate enhancement by the Kozai-Lidov effect should be less important. Thus, we do not include this effect in our calculation, which needs further study in our future work. Note that we also neglect the influence of resonant relaxation since it has no significant influence on the time average TDE rate\,\citep[][]{Rauch1996}.

For the different TDE overenhancement mechanisms, more observational tests are preferred to distinguish them. For instance, in the cases that the TDE host galaxies exhibit signs of SMBH binary activity and/or evidence of kicked SMBH, it may indicate a significant influence of SMBH binary on TDEs. For other secular dynamical effects, such as radial anisotropies, nuclear triaxiality and eccentric nuclear stellar disk, the TDE host galaxies should be distinguishable based on their morphology. Our model is similar to the model of overdense central star clusters, which could be proved by future higher-resolution imaging. Indeed, high-resolution HST images have shown that the TDE host galaxies have higher central concentrations on pc-kpc scale\,\citep[][]{Law-Smith2017ApJ,Graur2018ApJ,French2020ApJ}. In addition, we calculated the TDE rates by considering possible star-disk interactions in both the active and non-active phases of galaxies. Our model suggests a significant enhancement in TDE rates during the turn-off of AGNs, coinciding with the disappearance of the standard thin disk as the accretion rate decreases. In this case, we connect the TDE rates with the evolution of galaxies and predict that a fraction of SMBHs will exhibit weak nuclear activities in TDE galaxies. It should be noted that this fraction may also suffer uncertainties in the timescale for transitioning from an active to a non-active state. In the currently observed TDEs, several candidates are situated within low-luminosity AGNs\,\citep[e.g.,][]{Blanchard2017ApJ,Yan2018MNRAS,Liu2020ApJ,Shi2022ApJ,Homan2023A&A}, which support our predictions. Further observations of TDEs and their host galaxies in the time-domain era can shed light on the relationship between SMBH activities and the evolution of the galactic nuclear environment. Furthermore, future high-sensitivity X-ray surveys targeting TDE host galaxies can reveal the nuclear activities of SMBHs directly.

\begin{acknowledgments} 
We thank the referee for the thorough and helpful suggestions that greatly improved the manuscript. M.W. thanks Xiao Fan, Jiancheng Wu and Xiangli Lei for helpful discussions.
Y.M. is supported by the university start-up fund provided by Huazhong University of Science and Technology. Q.W. is supported by the National SKA Program of China (2022SKA0120101), the National Natural Science Foundation of China (grants U1931203 and  12233007) and the science research grants from the China Manned Space Project (No. CMS-CSST-2021-A06). The authors acknowledge Beijing PARATERA Tech CO., Ltd. for providing HPC resources that have contributed to the research results reported within this paper.
\end{acknowledgments}

\bibliography{ref}{}
\bibliographystyle{aasjournal}

\end{document}